\documentclass[epj]{webofc}
\usepackage[varg]{txfonts}   
\usepackage{graphicx,color} 
\usepackage{bm}       
\usepackage{amsmath}  
\usepackage{amssymb}
\usepackage{makeidx}
\usepackage{amsfonts}
\usepackage{rotating}
\usepackage{epsfig}

\woctitle{The XXIII International Workshop High Energy Physics and Quantum Field Theory}

\begin{document}
\title{The correction of hadronic nucleus polarizability
to hyperfine structure of light muonic atoms}

\author{\firstname{A.V.} \lastname{Eskin}\inst{1}\fnsep\thanks{\email{eskin33@mail.ru}} 
\and
\firstname{A.P.} \lastname{Martynenko}\inst{1}\fnsep\thanks{\email{a.p.martynenko@samsu.ru}}
\and
\firstname{E.N.} \lastname{Elekina}\inst{2}}

\institute{Samara U., 443086, Samara, Moskovskoe shosse 34
\and
Samara State U. of Architecture and Civil Engineering, 443001, Samara, Molodogvardeiskaya Str. 194
}

\abstract{The calculation of hadronic polarizability contribution of the nucleus to hyperfine structure 
of muonic hydrogen and helium is carried out within the unitary isobar model and experimental 
data on the polarized structure functions of deep inelastic lepton-proton and lepton-deuteron scattering. 
The calculation of virtual absorption cross sections of transversely and longitudinally 
polarized photons by nucleons in the resonance region is performed in the framework of the program MAID.
}
\maketitle
\section{Introduction}
\label{intro}

Precise investigation of the Lamb shift and hyperfine structure
of light muonic atoms is a fundamental problem for testing the Standard model and establishing 
the exact values of its parameters, as well as searching for effects of new physics. 
At present, the relevance of these studies is primarily related to
experiments conducted by the collaboration
CREMA (Charge Radius Experiments with Muonic Atoms) 
\cite{crema1,crema2,crema3,crema4} with muonic hydrogen and deuterium by methods
of laser spectroscopy.
So, as a result of measuring the transition frequency $ 2P^{F=2}_{3/2}-2S^{F=1}_{1/2}$
a more accurate value of the proton charge radius was found to be $r_E = 0.84087(39)$ fm,
which is different from the value recommended by CODATA for $7\sigma$ 
\cite{mohr}. The CODATA value is based
on the spectroscopy of the electronic hydrogen atom and on electron-nucleon
scattering. The measurement of the transition frequency 
$2P^{F=1}_{3/2}-2S^{F=0}_{1/2}$ for the singlet
$2S$ of the state $ (\mu p)$ allowed to obtain the hyperfine splitting of the $2S$ energy level
in muonic hydrogen, and also the values of the Zemach's radius $r_Z=1.082(37)$ fm and magnetic
radius $r_M=0.87(6)$ fm. The first measurement of three transition 
frequencies between energy levels $2P$ and $2S $ for
muonic deuterium $(2S_{1/2}^{F=3/2}-2P_{3/2}^{F=5/2})$, $(2S_{1/2}^{F =1/2}-2P_{3/2}^{F =3/2})$, 
$ (2S_{1/2}^{F=1/2}-2P_{3/2}^{F=1/2})$
allowed to obtain in 2.7 times the more accurate value of the charge radius of the deuteron,
which is also less than the value recommended by CODATA \cite{mohr}, by $7.5 \sigma$ 
\cite{crema4}. 
As a result, a situation emerges when there is an inexplicable discrepancy between the values 
of such fundamental
parameters, like the charge radius of a proton and deuteron, obtained from electronic and muonic
atoms. In the process of searching for possible solutions of the proton charge radius "puzzle"
various hypotheses were formulated, including the idea of the nonuniversality of the interaction 
of electrons and muons with nucleons. Preliminary experimental
data for muonic helium ions show that there is no large discrepancy in obtained
charge radii in comparison with CODATA.

In the experiments of the CREMA collaboration one very important task is solved: 
to obtain an order of magnitude more accurate values of the charge radii of the 
simplest nuclei (proton, deuteron, helion, alpha particle ....) that enter 
into one form or another into theoretical expressions for intervals of fine 
or hyperfine structure of the spectrum.
In this case, high sensitivity of characteristics of the bound muon to
distribution of charge density and magnetic moment of the nucleus is used.
Successful realization of this program is possible only in combination with precise theoretical
calculations of various energy intervals, measured experimentally. In this way,
the problem of a more accurate theoretical construction of the particle interaction operator in
quantum electrodynamics, the calculation of new corrections in the energy spectrum of muonic atoms
acquires a special urgency \cite{paper1,paper2,paper3}. 
The aim of this work consists in the calculation of the deuteron, helion and triton
polarizability correction to the hyperfine splitting (HFS). We perform a
calculation of hadronic polarizability  contribution using the isobar model describing the processes
of photo- and electroproduction of $\pi$, $\eta$ mesons, nucleon resonances
on the nucleon in the resonance region, and experimental data on the nucleon and deuteron 
polarized structure functions obtained in non-resonance region.

\section{General formalism}
\label{sec-1}

\begin{figure}[htbp]
\centering
\includegraphics[scale=0.7]{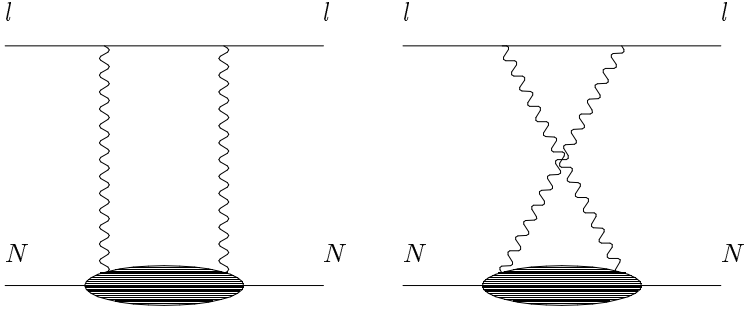}
\caption{Two-photon Feynman amplitudes determining the correction of the
nucleus polarizability to the hyperfine splitting of muonic atom.}
\label{fig-1}
\end{figure}

The leading order polarizability contribution to HFS is determined by two-photon exchange diagrams,
shown in Fig.~\ref{fig-1}. The corresponding amplitudes of virtual Compton
scattering on the nucleus can be represented as a convolution of antisymmetric parts of the lepton 
and hadron tensors which have the following form \cite{Z,cfm}:
\begin{equation}
\label{eq:1}
L_{\mu\nu}^{A}=\frac{1}{4}Tr\Bigl\{(1+\gamma^0)\gamma_5\hat s_1\Bigl[\gamma_1^\mu \frac{\hat p_1+\hat k+m_1}
{(p_1+k)^2-m_1^2}\gamma_1^\nu+\gamma_1^\nu \frac{\hat p_1-\hat k+m_1}
{(p_1-k)^2-m_1^2}\gamma_1^\mu\Bigr]\Bigr\},
\end{equation}
\begin{equation}
\label{eq:2}
W_{\mu\nu}^{A}=i\epsilon_{\mu\nu\alpha\beta}k^\alpha\Bigl\{s_2^\beta \frac{H_1(\nu,Q^2)}{(p_2\cdot k)}+
\frac{[(p_2k)s_2^\beta-(s_2k)p_2^\beta]}{(p_2\cdot k)^2}H_2(\nu,Q^2)\Bigr\},
\end{equation}
where $m_1$, $m_2$ are the lepton and nucleus masses, the nucleus four-momentum $p_2=(m_2,0)$,
$\epsilon_{\mu\nu\alpha\beta}$ is the totally antisymmetric tensor in four dimensions.
$s_1$, $s_2$ are spin four vectors of the lepton and nucleus. $H_1$, $H_2$ are the structure
functions of polarized scattering. The invariant quantity $p_2\cdot k$ is related to the energy
transfer $\nu$ in the nucleus rest frame: $p_2\cdot k=m_2 \nu$. The invariant mass of the
electroproduced hadronic system $W$ is then $W^2=m_2^2+2m_2\nu-Q^2=m_2^2+Q^2(1/x-1)$.
After computing the convolution of two tensors \eqref{eq:1} and \eqref{eq:2}, we obtain:
\begin{equation}
\label{eq:3}
L^{A}_{\mu\nu}W^{A}_{\mu\nu}=\frac{4}{3}({\bf s}_1{\bf s}_2)\frac{m_2k^2}{k^4-4m_1^2k_0^2}
\left[(k_0^2+2k^2)\frac{H_1}{(p_2\cdot k)}+3k_0^2k^2\frac{H_2}{(p_2\cdot k)^2}\right].
\end{equation}
According to the optical theorem the imaginary part of the forward Compton amplitude is related to
the cross section of inelastic scattering of off-shell photons from protons: $Im H_1(\nu,Q^2)=g_1(\nu,Q^2)/\nu$,
$Im H_2(\nu,Q^2)=m_2g_2(\nu,Q^2)/\nu^2$. As a result, neglecting the lepton mass the nucleus 
polarizability contribution to HFS can be presented in the form \cite{mf2002,mf2002a,carlson1,carlson}:
\begin{equation}
\label{eq:4}
\Delta E^{hfs}_{pol}=\frac{Z\alpha m_1}{2\pi m_2\mu_N}E_F(\Delta_1+\Delta_2)=
(\delta_1^p+\delta_2^p)E_F=\delta_{pol}E_F,
\end{equation}
\begin{equation}
\label{eq:5}
\Delta_1=\int_0^\infty\frac{dQ^2}{Q^2}\Bigl\{\frac{9}{4}F_2^2(Q^2)-
4m_2\int_{\nu_{th}}^\infty\frac{d\nu}{\nu^2}\beta_1(\theta)g_1(\nu,Q^2)\Bigr\},
\end{equation}
\begin{equation}
\label{eq:6}
\Delta_2=-12m_2\int_0^\infty\frac{dQ^2}{Q^2}
\int_{\nu_{th}}^\infty \frac{d\nu}{\nu^2}\beta_2(\theta)g_2(\nu,Q^2),
\end{equation}
where $\nu_{th}$ determines the pion-nucleus threshold:
\begin{equation}
\label{eq:7}
\nu_{th}=m_\pi+\frac{m_\pi^2+Q^2}{2m_2},
\end{equation}
and the functions $\beta_{1,2}$ have the form:
\begin{equation}
\label{eq:8}
\beta_1(\theta)=3\theta-2\theta^2-2(2-\theta)\sqrt{\theta(\theta+1)},
\end{equation}
\begin{equation}
\label{eq:9}
\beta_2(\theta)=1+2\theta-2\sqrt{\theta(\theta+1)},~\theta=\nu^2/Q^2.
\end{equation}
$F_2(Q^2)$ is the Pauli form factor of the nucleus.
The dependence on the mass of the lepton in \eqref{eq:5}-\eqref{eq:6} can also be taken into account, 
which leads to a certain modification of the functions $\beta_i(\nu,Q^2)$ \cite{cfm,carlson}.
This is important for increasing the accuracy of calculations in the case of muonic atoms.

Having information on the polarization structure functions of the nuclei, one can integrate into \eqref{eq:4} 
and obtain this correction.
The nucleus spin-dependent structure functions $g_1(\nu,Q^2)$, $g_2(\nu,Q^2)$
can be measured in the inelastic scattering of polarized leptons on polarized nuclei.
Accurate measurements of proton and deuteron polarized structure functions were made
in SLAC, CERN and DESY \cite{Abe1,Abe2,Anthony,Mitchell,Adams,Adeva}. These experimental data
can be used for the calculation the contribution in the nonresonance region where the invariant mass
$W$ must be greater than the mass of any resonance $N^\ast$ in the reaction $\gamma^\ast+N\to N^\ast$.
The threshold between the resonance region and the deep-inelastic region is not well defined, but it
is usually taken to be at about $W^2=4~GeV^2$. On the other hand in the resonance region we need 
theoretical model describing polarized structure functions $g_{1,2}(\nu,Q^2)$ 
since experimental data in this region are clearly insufficient. First of all, we can express functions 
$g_{1,2}(\nu,Q^2)$ in terms of virtual photon absorption cross sections as follows:
\begin{equation}
\label{eq:10}
g_1(\nu,Q^2)=\frac{m_2\cdot K}{8\pi^2\alpha(1+Q^2/\nu^2)}\left[\sigma^T_{1/2}
(\nu,Q^2)-\sigma^T_{3/2}(\nu,Q^2)+\frac{2\sqrt{Q^2}}{\nu}\sigma^{TL}_{1/2}(\nu,Q^2)\right],
\end{equation}
\begin{equation}
\label{eq:11}
g_2(\nu,Q^2)=\frac{m_2\cdot K}{8\pi^2\alpha(1+Q^2/\nu^2)}\left[-\sigma^T_{1/2}
(\nu,Q^2)+\sigma^T_{3/2}(\nu,Q^2)+\frac{2\nu}{\sqrt{Q^2}}\sigma^{TL}_{1/2}(\nu,Q^2)\right],
\end{equation}
where $K$ is the kinematical flux factor for virtual photons. The virtual photon absorption
cross sections have superscripts referring to the initial and final photon polarization being
longitudinal L or transverse T. The superscript TL is for the case when the photon polarization
direction changes during the interaction. The subscripts refer to the total spin of the photon-nucleus system.

Let us briefly describe the basic formulas that underlie the numerical results.
To construct the polarized structure functions \eqref{eq:10}-\eqref{eq:11} in resonance region we use the
Breit-Wigner parameterization for the photoabsorption cross sections 
\cite{Walker,Arndt,Teis1,Teis2,Krusche,Bianchi,D,Dong}. In the considered region of the variables 
$k^2$, $W$ the most important contribution is given by five resonances: $P_{33} (1232)$,
$S_{11} (1535)$, $D_{13} (1520)$, $P_{11} (1440)$, $F_{15} (1680)$. Accounting for the resonance decays to the $N\pi-$ and
$N\eta-$ states we can express the absorption cross sections
$\sigma^T_{1/2}$ and $\sigma^T_{3/2}$ as follows:
\begin{equation}
\label{eq:12}
\sigma^T_{1/2,3/2}=\left(\frac{k_R}{k}\right)^2\frac{W^2\Gamma_\gamma\Gamma_{R
\rightarrow N\pi}}{(W^2-M_R^2)^2+W^2\Gamma_{tot}^2}\frac{4m_N}{M_R\Gamma_R}
|A_{1/2,3/2}|^2,
\end{equation}
where $A_{1/2,3/2}$ are transverse electromagnetic helicity amplitudes,
\begin{equation}
\label{eq:13}
\Gamma_\gamma=\Gamma_R\left(\frac{k}{k_R}\right)^{j_1}\left(\frac{k_R^2+X^2}
{k^2+X^2}\right)^{j_2},~~X=0.3~GeV.
\end{equation}
The resonance parameters $\Gamma_R$, $M_R$, $j_1$, $j_2$, $\Gamma_{tot}$ are taken from \cite{PDG,Teis1,Teis2}. 
In accordance with \cite{Teis1,Krusche} the parameterization
of one-pion decay width is
\begin{equation}
\label{eq:14}
\Gamma_{R\rightarrow N\pi}(q)=\Gamma_R\frac{M_R}{M}\left(\frac{q}{q_R}\right)^3
\left(\frac{q_R^2+C^2}{q^2+C^2}\right)^2,~C=0.3~GeV
\end{equation}
for the $P_{33}(1232)$ and
\begin{equation}
\label{eq:15}
\Gamma_{R\rightarrow N\pi}(q)=\Gamma_R\left(\frac{q}{q_R}\right)^{2l+1}
\left(\frac{q_R^2+\delta^2}{q^2+\delta^2}\right)^{l+1},
\end{equation}
for resonances $D_{13}(1520)$, $P_{11}(1440)$, $F_{15}(1680)$. $l$ is the pion angular
momentum and $\delta^2$ = $(M_R-$ $m_N-m_\pi)^2$ + $\Gamma_R^2/4$. Here $q$
$(k)$ and $q_R$ $(k_R)$ denote the c.m.s. pion (photon) momenta of resonances
with the mass M and $M_R$ respectively. In the case of $S_{11}(1535)$ we take
into account $\pi N$ and $\eta N$ decay modes \cite{Krusche}:
\begin{equation}
\label{eq:16}
\Gamma_{R\rightarrow\pi,\eta}=\frac{q_{\pi,\eta}}{q}b_{\pi,\eta}\Gamma_R
\frac{q_{\pi\eta}^2+C_{\pi,\eta}^2}{q^2+C_{\pi,\eta}^2},
\end{equation}
where $b_{\pi,\eta}$ is the $\pi$ ($\eta$) branching ratio.

\begin{figure}[htbp]
\centering
\includegraphics[scale=0.5]{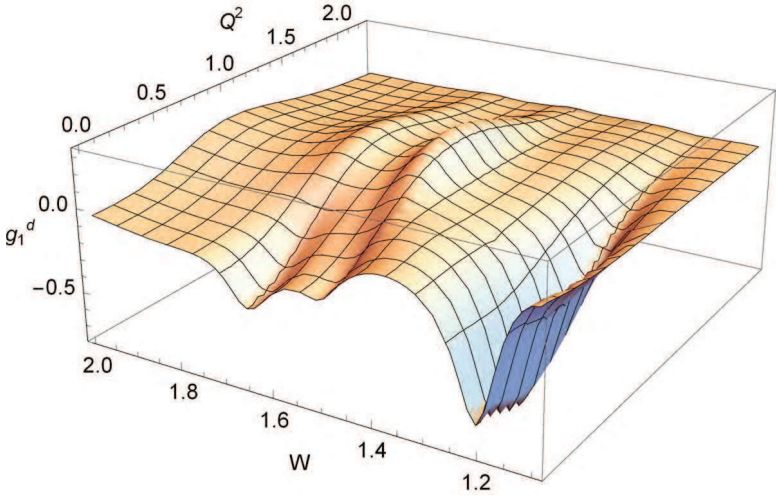}
\caption{The deuteron polarized structure function $g_1^d(W,Q^2)$ as function of $Q^2$ $(0\div 2.0.~GeV^2$
and W $(1.1\div 2.0)~GeV$.}
\label{fig-2}
\end{figure}

\begin{figure}[htbp]
\centering
\includegraphics[scale=0.45]{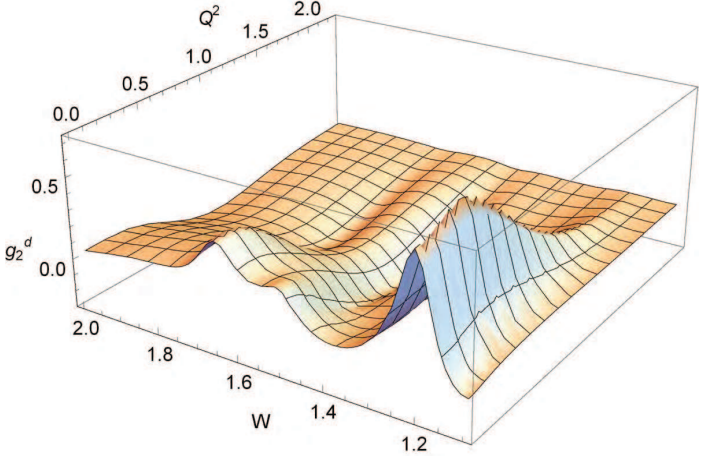}
\caption{The deuteron polarized structure function $g_2^d(W,Q^2)$ as function of $Q^2$ $(0\div 2.0.~GeV^2$
and W $(1.1\div 2.0)~GeV$.}
\label{fig-3}
\end{figure}

The cross section $\sigma^{TL}_{1/2}$ is determined by an expression similar to \eqref{eq:12}, 
containing the product $(S^\ast_{1/2}\cdot A_{1/2}+A_{1/2}^\ast S_{1/2})$ \cite{Abe1}. The calculation of
helicity amplitudes $A_{1/2}$, $A_{3/2}$ and longitudinal amplitude $S_{1/2}$, as functions of $Q^2$, was done on the basis
of constituent quark model (CQM) in \cite{Dong2,Isgur,CL,Capstick,LBL,Warns}.
The program of numerical calculation of cross sections \eqref{eq:12} was successfully realized by the authors of \cite{maid1,maid2}
within the unitary isobar model framework known as the MAID package (http://www.kph-uni-mainz.de/MAID).
In the unitary isobar model \cite{maid1,maid2} accounting for the Born terms,
the vector meson, nucleon resonance contributions and interference terms we calculate the cross sections
$\sigma^{T}_{1/2,3/2}$, $\sigma^{TL}_{1/2}$ by means of numerical program MAID in the resonance region as the
functions of two variables $W$ and $Q^2$. The obtained nucleon polarized structure functions $g_{1,2}(W,Q^2)$ are then used
for a construction of nucleus structure functions and calculation the polarizability contribution.
In Figs.~\ref{fig-2},\ref{fig-3} we show the obtained structure functions $g_{1,2}(W,Q^2)$.

Our calculation of polarizability contribution in deep inelastic region is based on experimental
data from \cite{Abe1,Abe2,Anthony,Mitchell,Adams,Adeva}. As was shown in previous paper \cite{cfm}
nucleon polarized structure functions can be expressed through polarized quark and gluon distributions
which obey to the evolution equations \cite{altarelli,hirai}. Solving $Q^2$ evolution equations we can construct 
nucleon functions $g_1(\nu,Q^2)$, $g_2(\nu,Q^2)$ which agree well with experimental data and the following
parameterization \cite{Abe1,Abe2,Anthony,Mitchell,Adams,Adeva,erbacher}:
\begin{equation}
\label{eq:17}
g_1^{p,d}(x,Q^2)=a_1x^{a_2}(1+a_3x+a_4x^2)[1+a_5f(Q^2)]F_1^{p,d}(x,Q^2),
\end{equation}
where the superscript index p,d corresponds to the proton or deuteron. Numerical integration is performed with $f(Q^2)=-\ln Q^2$
corresponding to the perturbative QCD behaviour. The calculation of the second part of the correction $\delta_{pol}$
in \eqref{eq:4}
in nonresonance region is carried out by means of Wandzura-Wilchek relation as in \cite{cfm}.

\section{Numerical results}
\label{sec-2}

Most part of numerical calculation is devoted to muonic deuterium.
To the hadronic contribution we include contributions that are determined by the nuclear reactions of the production 
of $\pi$-, $\eta$, and $K$-mesons on nucleons, the production of nucleon resonances.
In the approximation, which is then used for the calculation
the deuteron appears as a loosely coupled system of the proton and neutron,
so the deuteron polarized structure function can be presented as a sum of the proton
and neutron structure functions:
\begin{equation}
\label{eq:18}
g_i^d(W,Q^2)=g_i^p(W,Q^2)+g_i^n(W,Q^2).
\end{equation} 
The MAID program allows us to calculate separately the proton and neutron structure functions $g_i^{p,n}(W,Q^2)$.

\begin{table}[h]
\centering
\caption{The contributions to the GDH sum rule for the proton and neutron.}
\label{tb1}
\begin{tabular}{|c|c|c|}   \hline
 &&\\
Contribution to GDH integral & $I_{GDH}^p$, $\mu b$  &  $I_{GDH}^n$, $\mu b$   \\ 
&&  \\  \hline
 &&\\
Contribution of $N\pi$ states  &  165.9      &   133.2  \\  
 &&\\ \hline
  &&\\
Contribution of $N\eta$ states  &   -8.9     &  -5.7   \\   
 &&\\ \hline
  &&\\
Contribution of K-mesons  &  -1.8   &   -3.0   \\   
 &&\\  \hline
  &&\\
Contribution of $ N\pi\pi$ states  & 47.8  & 50.5  \\    
 &&\\  \hline
  &&\\
Total contribution  & 203.0  & 175.0  \\    
 &&\\  \hline
  &&\\
The GDH value   & 204.8   & 232.5   \\   
 &&\\  \hline
\end{tabular}
\end{table}

For the integration in \eqref{eq:4} an important role is played by the Gerasimov-Drell-Hern (GDH) sum rule 
\cite{sbg,dh,krein,costanza,aa} which connects an energy-weighted integral of the difference of the helicity dependent
real-photon absorption cross sections with the anomalous contribution $\kappa=\frac{\mu_N m_2}{es_2}-Z$
to the magnetic moment $\mu$ of the nucleus:
\begin{equation}
\label{eq:19}
I_{GDH}=\int_{\nu_{th}}^\infty\frac{\sigma_p-\sigma_a}{\nu}d\nu=4\pi^2\kappa^2\frac{\alpha s_2}{m_2^2},
\end{equation}
where $\nu$ is the photon energy, $\sigma_p$ and $\sigma_a$ are the total photoabsorption cross
sections for parallel and antiparallel orientation of photon and nucleus spins, respectively.
The lower limit of the integral, $\nu_{th}$, corresponds to pion production and photodisintegration 
threshold for a nucleonic and nuclear target, respectively.
Strictly speaking, in order for the integral in \eqref{eq:4} to converge over the variable $Q$, the sum rule 
\eqref{eq:19} must exactly be satisfied.
The GDH sum rule is satisfied for any nucleus, including a proton and a neutron. Since the deuteron 
structure function satisfies \eqref{eq:18}, it is necessary to achieve the sum rules for the proton and neutron.  
Since the neutron and the proton have large anomalous magnetic moments $\kappa_p=1.79$, $\kappa_n=-1.91$), we obtain large values 
for the integral $I^{p,n}_{GDH}$ \eqref{eq:19}. In turn, in the case of a deuteron, a small value 
of AMM ($\kappa_d=-0.143$) leads to a small value $I^{d}_{GDH}=-0.65~\mu b$ \eqref{eq:19}, 
which is two orders of magnitude smaller than the 
corresponding values for nucleons. 
When the deuteron is represented in the form of a state of two almost 
free nucleons, we find that the quantity in right part of \eqref{eq:19} is equal to 437 $\mu b$. 
There are also other channels ($\gamma d\to pn$) which can not be
treated in the quasi-free approximation but which contribute to \eqref{eq:19}.
An analysis carried out in \cite{ha1,ha2} 
showed that this value decreases significantly when taking into account the photodisintegration channel, 
which is not considered in this paper.
Therefore, representing the deuteron in the form of the sum of a proton and a neutron, we add in \eqref{eq:4} 
two terms with the Pauli form factors of the proton and the neutron
to ensure the fulfilment of the sum rule and the subsequent integration into \eqref{eq:4}.
Calculations in the MAID show that the sum rule 
for a proton is satisfied with a sufficiently high accuracy, whereas for a neutron the difference between the left 
and right parts in \eqref{eq:19} reaches $30\%$ (see Table~\ref{tb1}).

To avoid this difficulty and obtain however an estimate of hadronic contribution to the
neutron and deuteron polarizability we introduce a cutoff of the momentum integral over Q in \eqref{eq:4} 
at some value $\kappa\approx 0.01$ GeV, supposing that the GDH sum rule for the neutron holds exactly
and the region of small Q $(0\div 0.01)$ GeV does not give essential contribution to general value of
correction \eqref{eq:4}. Similar cutoff procedure at small values Q was used in \cite{khmil} with $\kappa\approx 0.045$
GeV, in which different corrections to deuterium hyperfine structure were considered from the two-photon
exchange amplitudes. As a result total value of the polarizability correction for muonic deuterium including
the resonance and nonresonance regions is equal to 0.13 meV. In the case of He-3 total polarizability
contribution 0.06 meV is determined by unpaired neutron because two protons have opposite spins and do not
contribute to hyperfine splitting. The similar situation occurs for the triton in which two neutrons form 
closed shell. In muonic tritium the polarizability correction is equal to 0.05 meV.
Total error of our calculation is estimated in $30~\%$ which is determined mainly by the
uncertainty from two-pion contribution and the cutoff procedure used above. The obtained
values of polarizability corrections should be used for obtaining total values of hyperfine splitting
in light muonic atoms \cite{apm2004,apm2008,paper1,apm2017}.

The work is supported by Russian Foundation for Basic Research (grant No. 16-02-00554).

\end{document}